# AI-interpreted Optical Scattering for Robust and Focal Depth-Aware Imaging

Eunji Ko[1], Patrick Ross[2], Corey Hart[2], Wolfgang Losert[1,*]

[1]University of Maryland – College Park
[2]Lockheed Martin

**ABSTRACT**. Optical scattering has conventionally been regarded as an impediment in imaging research due to the degradation of image quality during reconstruction. Nevertheless, this study explores two cases in which optical scattering may serve a beneficial role in image reconstruction tasks. We compared the No Scattering MNIST dataset with three Scattering MNIST datasets, each generated under distinct scattering conditions. To assess the information content of the resulting speckle patterns, we employed a Variational Autoencoder (VAE) approach which achieves accuracy comparable to state-of-the-art deep learning approaches, but has an interpretable latent space. We find that scattering can enhance data robustness against spatial pixel loss by effectively distributing information. We also demonstrate that scattering can enable distinctions of focal depth information. We anticipate that these findings will contribute to more efficient imaging techniques, particularly in the presence of obstacles and three-dimensional signals.

## I. INTRODUCTION.

Imaging through scattering media is a challenging problem across a broad range of applications, from fluorescence microscopy in turbid biological tissue to depth-resolved sensing in aerosols and colloidal suspensions. In these settings, multiple scattering events degrade spatial coherence and the signal captured by detectors. In many previous studies, scattering has therefore been treated as noise to be corrected, either by measuring the transmission matrix of the scattering process to characterize the full input–output field relationship [1, 2], or by training deep neural networks to invert the scattering transformation with U-Net or transformer-based architectures [3–6]. These deep learning models, trained on paired speckle–object datasets, can learn effective end-to-end mappings without explicit physical modeling of the scattering process. While these approaches reconstruct the original information effectively, they offer no mechanism for scattering to assist reconstruction or classification tasks.

Variational Autoencoders (VAEs) [7, 8] offer a generative alternative that addresses this limitation. By mapping observations to a structured probabilistic latent space, VAEs learn compact representations that capture the underlying generative factors of the data rather than merely fitting their parameters to a particular dataset. Unlike discriminative models, the VAE latent space affords a degree of interpretability: individual latent dimensions can be traversed and inspected to reveal which physical factors of variation they encode [9, 10], making VAEs particularly well-suited to problems where understanding the generative structure of the data is as important as predictive accuracy.

Two fundamental questions arise when applying such models to speckle imaging. The first is practical: how much of the detector plane is needed? Scattering redistributes light across the plane, but it is not known whether object-relevant information is concentrated near the optical axis or spread across the entire speckle field and how this depends on scattering strength. The second question concerns whether the scattering process itself encodes physically meaningful information beyond the object. Specifically, whether the focal depth of the object within the scattering volume leaves a decodable signature in the speckle pattern. Prior work has established that depth information survives scattering and can be recovered from a single speckle pattern [11, 12], suggesting that the speckle field simultaneously carries both object content and depth-related information. However, no existing framework has attempted to jointly decode and explicitly separate these two contributions within a single learned latent representation.

*Contact author: wlosert@umd.edu

In this work, we address both questions with VAE frameworks. We first apply circular occlusion masks of increasing radius to the speckle input and track reconstruction performance as a function of mask radius. Performance degrades in different ways for No -Scattering MNIST and Scattering MNIST datasets with increasing mask size, indicating that object information is encoded across the speckle field when scattered. In the second part, we present a split-latent VAE architecture, inspired by Domain-invariant VAE [13], that partitions the latent space into two disjoint subspaces: a 'digit' subspace capturing structural features, and a 'focal' subspace encoding depth-discriminative information reinforced by a jointly trained classifier head. The model reconstructs clean and scattered image estimates and the focal depth information from the latent space without requiring separate networks for reconstruction and depth estimation.

## II. METHODS

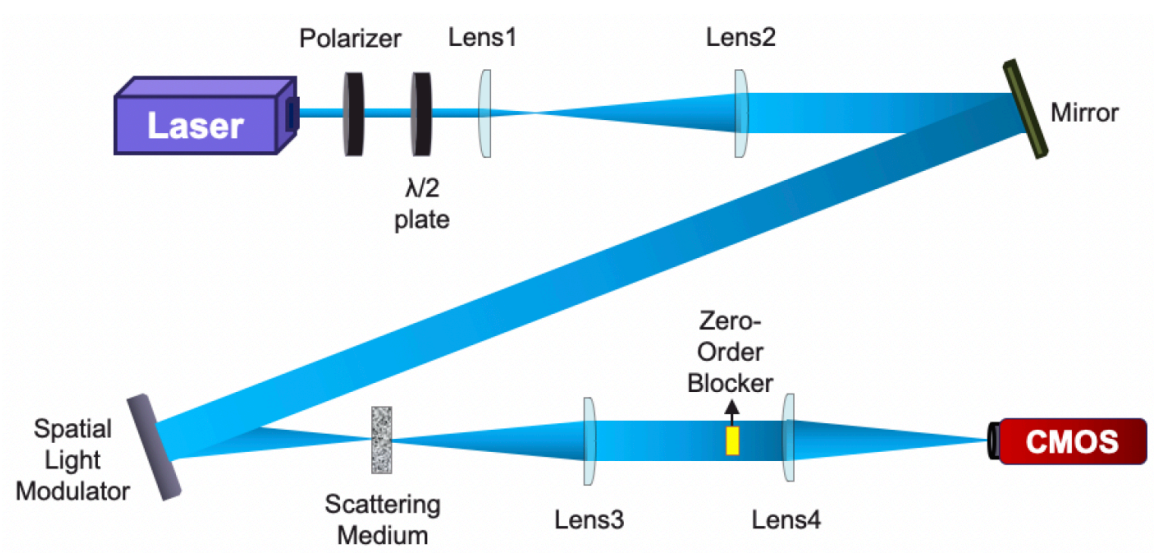

Figure 1. Optical scattering imaging setup. A linearly polarized, telescope-expanded laser beam illuminates an SLM, and the modulated output is imaged through a scattering medium onto a CMOS camera via a 4f relay with zero-order blocking at the Fourier plane between lens 3 and lens 4

### A. Experimental setup

Speckle patterns were generated by passing spatially modulated images, formed using a phase-only SLM (Holoeye LCR-2500, 1024 × 768 resolution, 19 µm pixel pitch), through a custom scattering medium, as shown in Fig 1. The input beam was provided by a collimated 473 nm laser (OBIS LX, 200 mW), expanded 20× to match the SLM aperture. Polarization was optimized using a polarizer and a half-wave plate to ensure efficient SLM modulation. The phase patterns were relayed through a 4f system. Unmodulated light was filtered at the Fourier plane, and the resulting speckle patterns were imaged onto a CMOS sensor (PointGrey, 1600 × 1200, 1/1.8"). A zero-order blocker is used in this setup with SLM to block the non-diffracted beam that passes straight through without modulation. This zero-order light often appears as a bright central spot and carries no useful information, therefore, it can obscure the modulated or diffracted patterns we want to analyze.

*Contact author: wlosert@umd.edu

The scattering medium was fabricated by pouring Polydimethylsiloxane (PDMS) into a 3D-printed mold. PDMS is commonly used as a scattering medium because it is optically transparent, biocompatible, and easy to mold or tune. We created three scattering media (low, moderate, high scattering) with varying scattering intensities by incorporating three distinct concentrations of zinc-oxide (ZnO) microparticles into PDMS. When the spatially structured light from the SLM passed through this medium, it generated complex speckle patterns that were employed for further analysis.

This study examines four distinct datasets. The first dataset, referred to as the 'No Scattering' MNIST dataset, was created by transmitting MNIST digit pattern generated by a spatial light modulator (SLM) directly to a camera, without the presence of a scattering medium. And there are three datasets labeled as 'Low Scattering', 'Moderate Scattering', and 'High Scattering' MNIST datasets. These were produced by transmitting the MNIST digit pattern generated by the SLM through a medium with different scattering strength.

### B. Computational methods

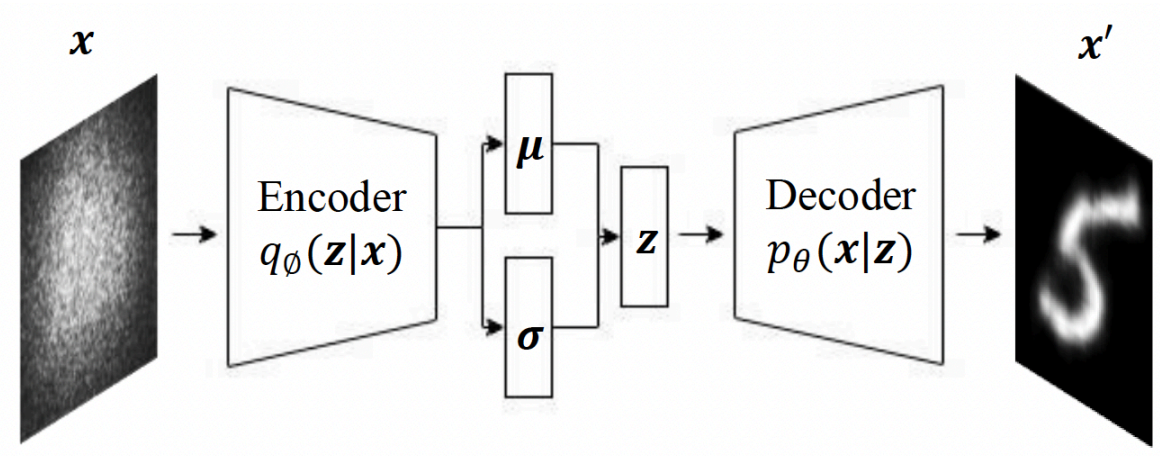

Figure 2. Illustration of variational autoencoder (VAE)

## 1. Variational autoencoder

We adopted a variational autoencoder (VAE) framework to reconstruct original information from the input images to target images. VAE is one of generative models that learns to encode input data into a probabilistic latent space and decode from the space back into output images.

### a. Standard VAE

As shown in Fig 2, given an observed image $\mathbf{x} \in \mathbb{R}^{H \times W}$, the VAE learns an approximate posterior distribution $q_\phi(\mathbf{z} \mid \mathbf{x})$ over latent variables $\mathbf{z}$, parameterized by an encoder network with parameters $\phi$. A decoder network with parameters $\theta$ then models the likelihood $p_\theta(\mathbf{x} \mid \mathbf{z})$. Training is performed by maximizing the Evidence Lower Bound (ELBO):

$$\mathcal{L}_{\text{ELBO}} = \mathbb{E}_{q_\phi(\mathbf{z}|\mathbf{x})}[\log p_\theta(\mathbf{x} \mid \mathbf{z})] - D_{\text{KL}}\big(q_\phi(\mathbf{z} \mid \mathbf{x}) \;\|\; p(\mathbf{z})\big)$$

where the first term is the reconstruction likelihood and the second term is the Kullback–Leibler (KL) divergence from the approximate posterior to the standard normal prior $p(\mathbf{z}) = \mathcal{N}(\mathbf{0}, \mathbf{I})$.

The encoder outputs the parameters of a diagonal Gaussian posterior:

$$q_\phi(\mathbf{z} \mid \mathbf{x}) = \mathcal{N}\big(\mathbf{z}; \boldsymbol{\mu}_\phi(\mathbf{x}), \text{diag}(\boldsymbol{\sigma}_\phi^2(\mathbf{x}))\big)$$

Latent samples are drawn via the reparameterization trick to enable gradient backpropagation:

$$\mathbf{z} = \boldsymbol{\mu}_\phi(\mathbf{x}) + \boldsymbol{\sigma}_\phi(\mathbf{x}) \odot \boldsymbol{\epsilon}, \qquad \boldsymbol{\epsilon} \sim \mathcal{N}(\mathbf{0}, \mathbf{I})$$

The KL divergence term for a diagonal Gaussian has a closed-form expression:

$$D_{\text{KL}}\big(q_\phi(\mathbf{z} \mid \mathbf{x}) \;\|\; p(\mathbf{z})\big) = -\frac{1}{2}\sum_{j=1}^{d}\big(1 + \log \sigma_j^2 - \mu_j^2 - \sigma_j^2\big)$$

where $d$ is the dimensionality of the latent space.

*Contact author: wlosert@umd.edu

### b. Split-Latent VAE

To disentangle the scattering-invariant image content from scattering-dependent focal depth information, we divide the latent space into two disjoint subspaces:

$$\mathbf{z} = [\mathbf{z}_{\text{digit}}, \mathbf{z}_{\text{focal}}]$$

where $\mathbf{z}_{\text{digit}} \in \mathbb{R}^{d_s}$ captures digit features common across focal depths, and $\mathbf{z}_{\text{focal}} \in \mathbb{R}^{d_f}$ encodes focal depth information that is blended with scattering information. The encoder produces a single concatenated posterior parameter vector $(\boldsymbol{\mu}, \log \boldsymbol{\sigma}^2) \in \mathbb{R}^{d_s + d_f}$, from which the subspace-specific parameters are obtained by index partitioning:

$$\boldsymbol{\mu}_s = \boldsymbol{\mu}_{[:d_s]}, \boldsymbol{\mu}_f = \boldsymbol{\mu}_{[d_s:]}$$

and analogously for $\log \boldsymbol{\sigma}_s^2$ and $\log \boldsymbol{\sigma}_f^2$. Latent samples are drawn via the reparameterization trick and split accordingly into $\mathbf{z}_{\text{digit}}$ and $\mathbf{z}_{\text{focal}}$.

The decoder is split into two separate paths with distinct inputs. The clean image reconstruction $\hat{\mathbf{x}}_{\text{clean}}$ is decoded exclusively from $\mathbf{z}_{\text{digit}}$, enforcing that the digit subspace alone must carry sufficient information to recover the digit content. The scattered image reconstruction $\hat{\mathbf{x}}_{\text{scat}}$ is decoded from the full latent code $\mathbf{z} = [\mathbf{z}_{\text{digit}}, \mathbf{z}_{\text{focal}}]$, allowing the focal subspace $\mathbf{z}_{\text{focal}}$ to encode also the scattering information needed to reproduce the observed speckle pattern, not only the coal depth information.

The full training loss consists of reconstruction, KL regularization, and focal classification terms:

$$\mathcal{L} = \mathcal{L}_{\text{recon}} + \beta \cdot \mathcal{L}_{\text{KL}} + \lambda \cdot \mathcal{L}_{\text{focal}}$$

The reconstruction loss is computed as the mean squared error between the model outputs and their respective targets:

$$\mathcal{L}_{\text{recon}} = \| \mathbf{x}_{\text{clean}} - \hat{\mathbf{x}}_{\text{clean}} \|_2^2 + \| \mathbf{x}_{\text{scat}} - \hat{\mathbf{x}}_{\text{scat}} \|_2^2$$

The KL loss is summed over both latent subspaces:

$$\mathcal{L}_{\text{KL}} = D_{\text{KL}}\big(q_\phi(\mathbf{z}_{\text{digit}} \mid \mathbf{x}) \;\|\; p(\mathbf{z})\big) + D_{\text{KL}}\big(q_\phi(\mathbf{z}_{\text{focal}} \mid \mathbf{x}) \;\|\; p(\mathbf{z})\big)$$

The focal depth classifier is a lightweight network $f_\psi$ that takes $\mathbf{z}_{\text{focal}}$ as input and predicts a probability distribution over $C$ discrete focal depth classes. The classification loss is the standard cross-entropy:

$$\mathcal{L}_{\text{focal}} = -\sum_{c=1}^{C} y_c \log \hat{p}_c, \hat{p}_c = \text{softmax}\big(f_\psi(\mathbf{z}_{\text{focal}})\big)_c$$

where $y_c \in \{0,1\}$ is the one-hot ground-truth focal depth label and $\hat{p}_c$ is the predicted probability for class $c$ . Overall, the hyperparameter $\beta$ controls the regularization strength of the KL term, and $\lambda$ weights the classification objective relative to reconstruction.

### 2. Computation details

To assess the semantic fidelity of the reconstructed images, we used a Stochastic Optimization of Plain Convolutional Neural Networks (SOPCNN) to classify them [14]. The SOPCNN was trained on original MNIST datasets and then tested on reconstructed outputs. This classification accuracy was used to quantify the preservation of class-discriminative features.

To mitigate dataset sampling bias, each experimental condition was run across five independently drawn random seeds, with all reported metrics reflecting test-set performance exclusively.

All models were implemented using PyTorch and trained on a workstation equipped with an NVIDIA GeForce RTX 4090 GPU with CUDA 12.3 support. GPU monitoring and performance management were conducted via NVIDIA-SMI.

*Contact author: wlosert@umd.edu

## III. RESULTS & DISCUSSION

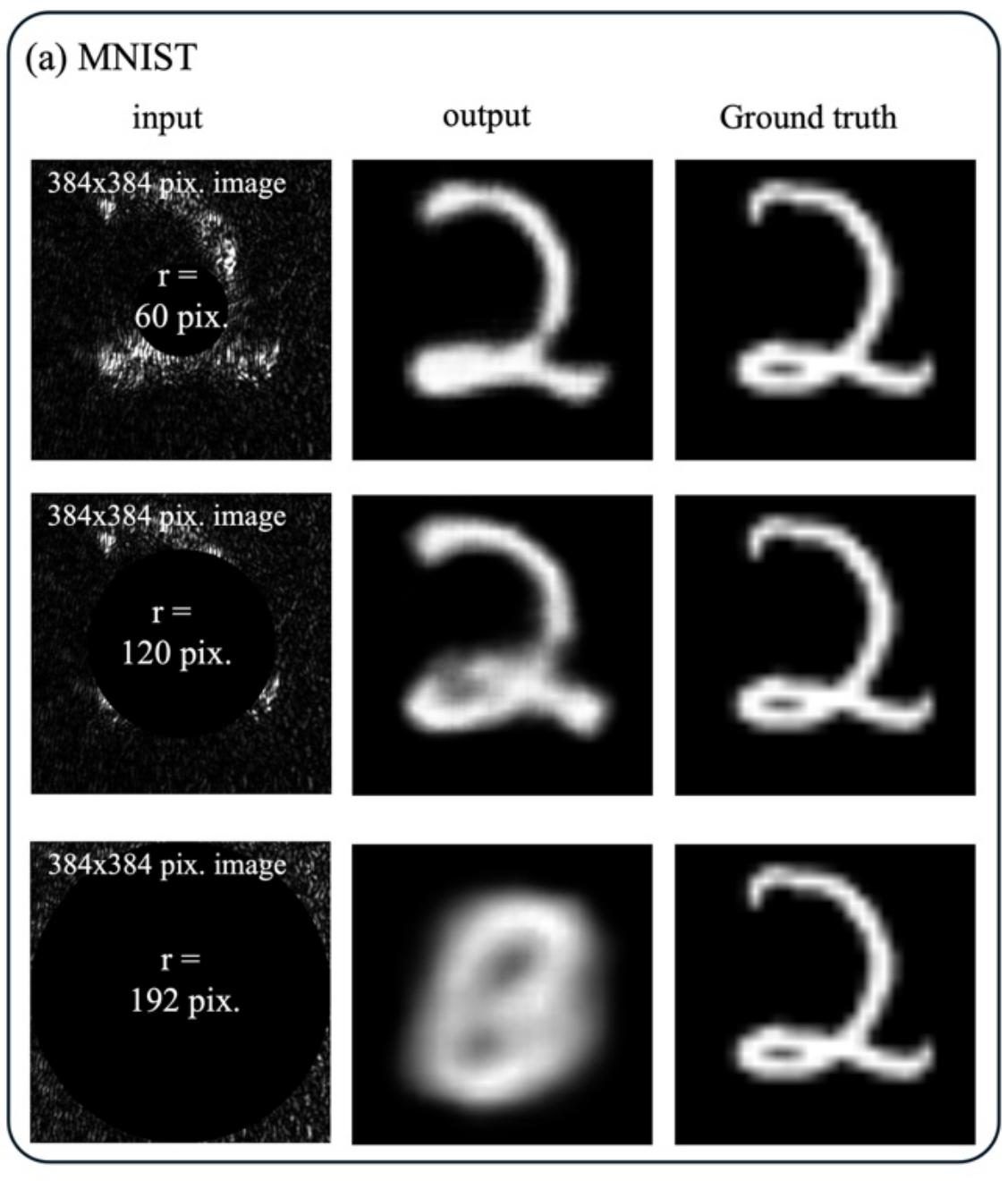

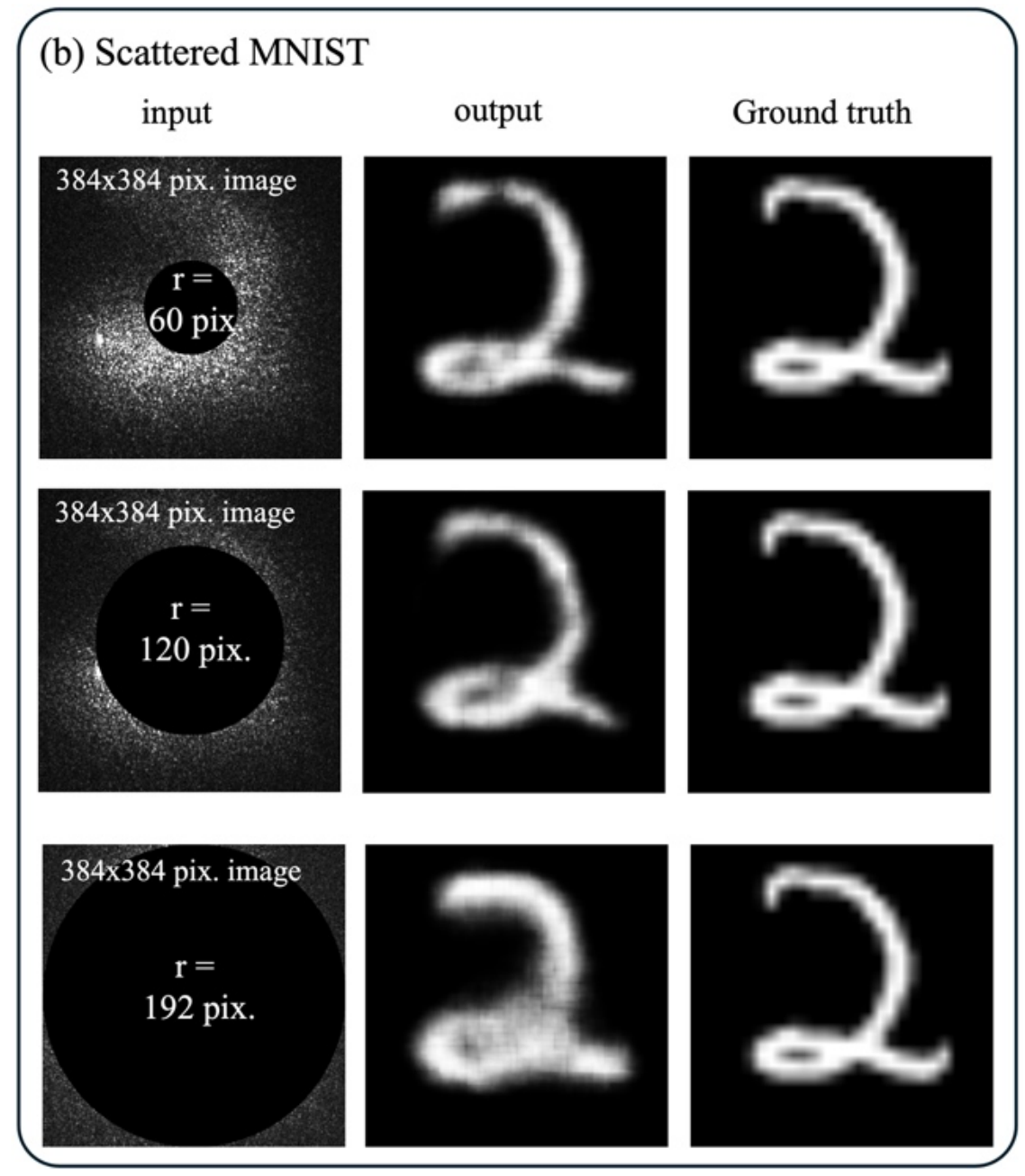

Figure 3. Visual examples of VAE reconstruction results with (a) the No-Scattering MNIST dataset and (b) the Moderate Scattering dataset. The first column shows the input, the second column displays the output, and the last column presents the ground truth (target). Each row represents the cases where a different radius of mask applied to the input images.

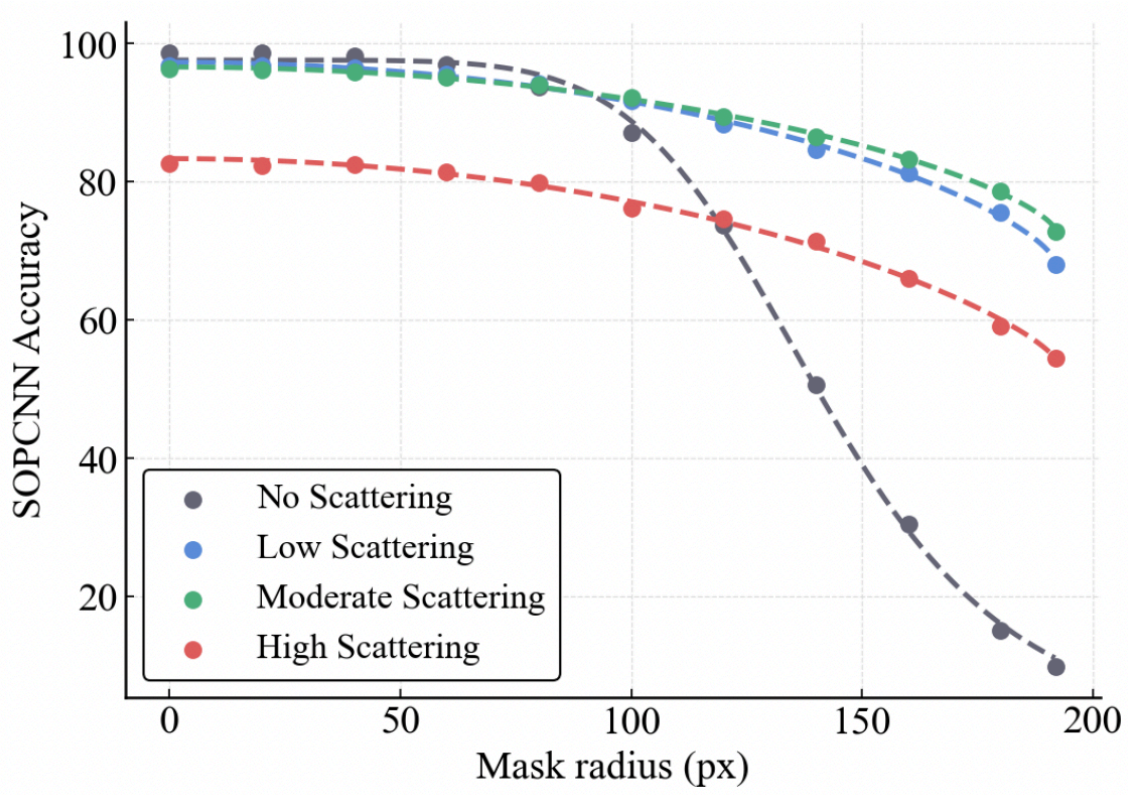

Fig 4. The reconstruction outcomes of center-occluded MNIST images using VAE, assessed by the accuracy of SOPCNN image classification, over increasing radius. No Scattering data was fitted with Sigmoid function, while Low/Moderate/High Scattering datasets were fitted with a power law function. All the data was fitted on average accuracies over N = 5 independent training seeds

Table I. Fitted power law parameter $\alpha$ for Scattering datasets. $\alpha \to 1$reflects increasingly uniform speckle statistics.

| | $\alpha$ |
|---|---|
| Low Scattering | 0.692 ± 0.030 |
| Moderate Scattering | 0.694 ± 0.022 |
| High Scattering | 0.757 ± 0.047 |

### A. Circular occlusion masking

In the initial segment of our results, we sought to determine whether the peripheral regions of images retain the original information when the central portion is obscured by a circular mask. Specifically, we applied a black circular mask, composed of zero-pixel intensities, to the center of the images. To assess the impact of increasing mask radius on image reconstruction, we varied the radius from 20 to 192 pixels, which is half the length of 384x384 images. Example of input, output, and target images for three different mask radii (60, 120, and 192) are presented in Fig. 3 (a)-(b). Notably, the reconstructed output images gradually lose detail and completely fail to reconstruct at a radius of 192 in the No Scattering MNIST dataset. Conversely, the output in the Moderate Scattering MNIST dataset retains its details as the mask radius increases, even when the images are nearly covered with a r=192 px. mask. The ability of the Scattered datasets to reconstruct the original images despite increasing radius is illustrated in greater detail in Fig. 4. In SOPCNN image classification task, the No Scattering MNIST dataset deteriorates rapidly as the mask radius increases, falling to approximately 10% in SOPCNN accuracy, indicative of a random guess. In contrast, the decline in the Scattering datasets is much less pronounced.

*Contact author: wlosert@umd.edu

To investigate how scattering redistributes spatial information analytically, we fitted each dataset to analytical functions that can describe each dataset's behavior.

For No Scattering MNIST, accuracy seems to follow a sigmoid decay, as shown in Fig 4:

$$A(r) = A_{\min} + \frac{A_{\max} - A_{\min}}{1 + \left(\frac{r}{r_c}\right)^k}$$

where $r_c$is the radius at which accuracy falls to the midpoint between its maximum and minimum values, and $k$ is the steepness coefficient describing how sharply accuracy collapses near $r_c$. The large (fitted) steepness coefficient $k = 6.69$ indicates that accuracy remains nearly unchanged for mask radii well below (fitted) $r_c$, then collapses rapidly over a narrow radius window centered at $r_c = 141.3$px, which turns out to be near the radius the mask reaches the boundary of digit stroke region.

For all Scattering datasets, a power law effectively characterizes their accuracy trends:

$$A(r) = A_{\min} + (A_{\max} - A_{\min})\left(1 - \frac{r^2}{R^2}\right)^{\alpha}$$

where $1 - r^2/R^2$is the unmasked area fraction and $\alpha$ governs how nonlinearly accuracy responds to visible area. Also, R=192 px. is the maximum mask radius applied, corresponding to the inscribed circle of the 384×384 size of image. When $\alpha = 1$, accuracy degrades in direct proportion to masked area, which means spatially uniform information distribution. And $\alpha$ increases monotonically with scattering density ($0.692 \to 0.757$), approaching the uniform limit $\alpha = 1$ at the heaviest scattering condition, which is consistent with heavier scattering producing more spatially homogeneous speckle statistics.

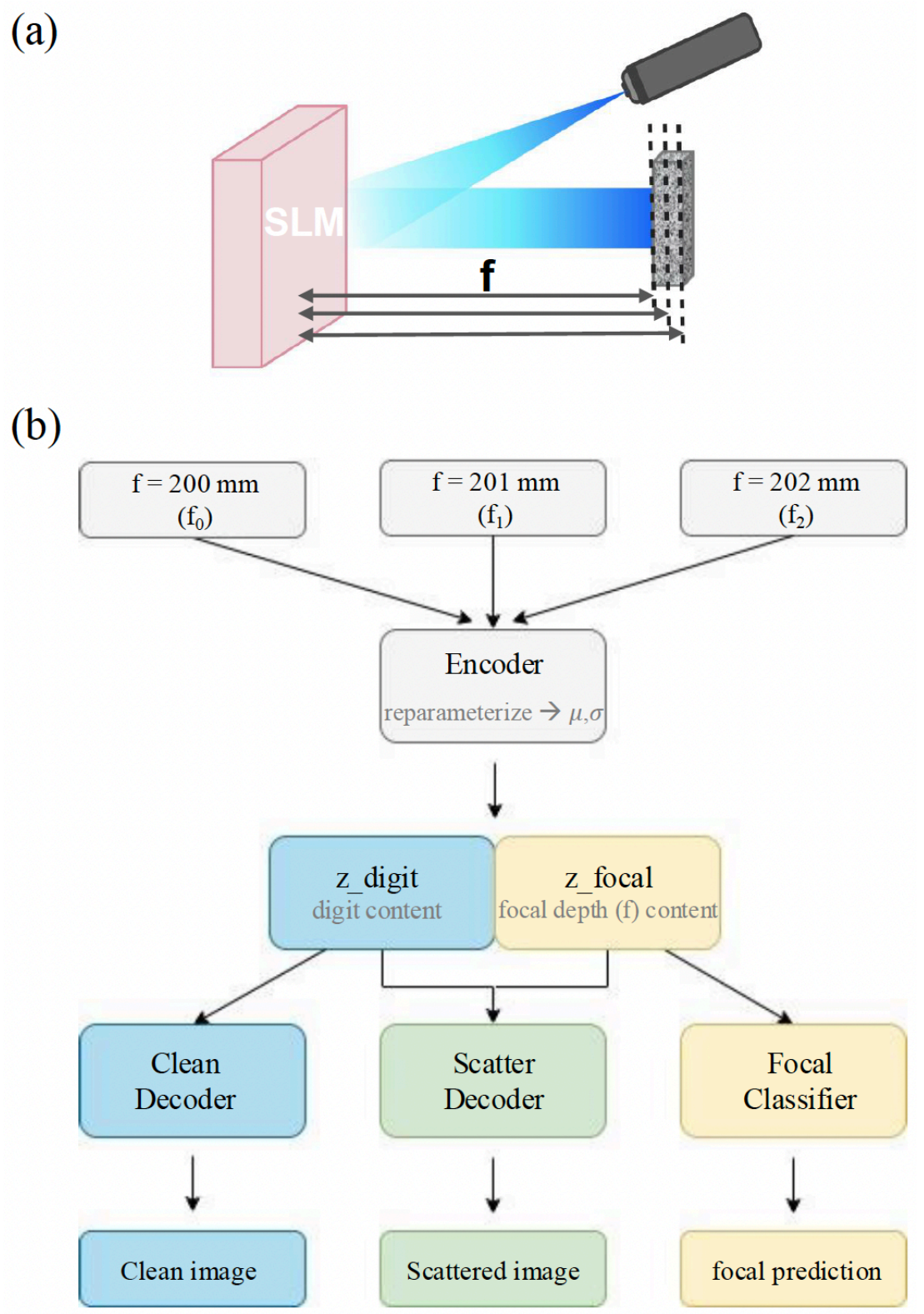

Fig 5. (a) The illustration of generating three different focal depth datasets (b) the schematic of the split-latent VAE we developed for the focal depth discrimination task

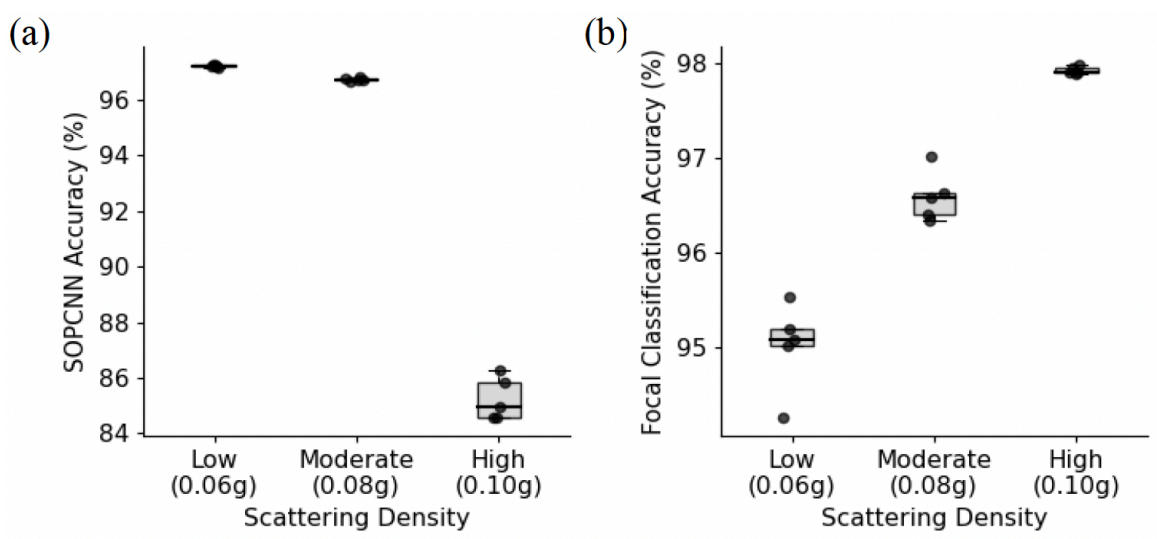

Fig 6. The outcomes of the split-latent VAE (a) SOPCNN accuracy for reconstructed clean MNIST images under different scattering conditions, and (b) focal classification accuracy on $\mathbf{z}_{\text{focal}} \in \mathbb{R}^6$ across different scattering scenarios. Each datapoints is from five different independent training seeds.

## B. Focal depth discrimination

In this section of our study, we explore whether optical scattering could effectively distinguish between the focal depths of three-dimensional scattering medium. Here, we projected the same MNIST pattern onto each of three focal depths of the scattering medium and collected speckle patterns for each depth separately as demonstrated in Fig 5(a). We then simultaneously input the three datasets into the encoder of split-latent VAE, as depicted in Fig 5(b). We designed the structure of this model such that the initial latent vector, $\mathbf{z}_{\text{digit}}$ is responsible solely for encoding MNIST digit data, while the second latent vector, $\mathbf{z}_{\text{focal}}$ is dedicated to encoding focal depth information. This is achieved by using only $\mathbf{z}_{\text{digit}}$ to decode into clear MNIST images and employing both $\mathbf{z}_{\text{digit}}$ and $\mathbf{z}_{\text{focal}}$ to decode into dispersed MNIST images. We deliberately chose not to include a third latent space for encoding scattering information, as we aimed to integrate this information specifically into $\mathbf{z}_{\text{focal}}$ to leverage the scattering effect. And $\mathbf{z}_{\text{focal}}$ is reinforced by a jointly trained classifier head. The SOPCNN accuracy of reconstructed images and focal classification accuracy on $\mathbf{z}_{\text{focal}}$ latent vector is shown in Fig 6(a) and (b), respectively. When scattering was low to moderate, the accuracy ranged from 96 to 97%, but with high scattering, the SOPCNN accuracy dropped to about 85%. Interestingly, even though high scattering had the lowest SOPCNN accuracy, it resulted in the highest accuracy for focal classification ~98%. Additionally, the accuracy of focal classification increased gradually as the scattering condition became more intense.

Table II. Silhouette score (SS) of $\mathbf{z}_{\text{focal}}$ across scattering densities, averaged over N = 5 independent training seeds (mean ± std).

| | SS |
|---|---|
| Low Scattering | 0.4237 ± 0.0230 |
| Moderate Scattering | 0.4420 ± 0.0140 |
| High Scattering | 0.5825 ± 0.0288 |

To assess focal depth discriminability in the latent space, we computed the silhouette score on $\mathbf{z}_{\text{focal}} \in \mathbb{R}^6$. For each scattering condition and each of the five training seeds, 1,000 samples per focal position were encoded using the posterior mean $\boldsymbol{\mu}$, yielding 3,000 points in $\mathbb{R}^6$ with known focal labels $\{f_0, f_1, f_2\}$. For each point $\mathbf{z}_n$ in cluster $C_i$, we compute $a(\mathbf{z}_n)$, the mean Euclidean distance to all other points in the same

*Contact author: wlosert@umd.edu

cluster, and $b(\mathbf{z}_n)$, the mean Euclidean distance to all points in the nearest other cluster:

$$a(\mathbf{z}_n) = \frac{1}{|C_i| - 1} \sum_{\mathbf{z}_m \in C_i, m \neq n} \| \mathbf{z}_n - \mathbf{z}_m \|_2,$$
$$b(\mathbf{z}_n) = \min_{j \neq i} \frac{1}{|C_j|} \sum_{\mathbf{z}_m \in C_j} \| \mathbf{z}_n - \mathbf{z}_m \|_2$$

The silhouette coefficient is $s(\mathbf{z}_n) = (b - a)/\max(a, b)$, and the silhouette score is the mean of $s$ across all points, ranging from $-1$ (point closer to another cluster than its own) to $+1$ (point well within its own cluster). As shown in Table II, the silhouette score remains lower in Low and Moderate Scattering then increases substantially in High Scattering, confirming that heavier scattering produces better-separated focal depth clusters in the latent space. The consistency of this effect across all five seeds indicates that this improvement reflects a property of the physical scattering regime rather than artifact from a particular training run.

Table III. Pairwise inter-cluster distances between focal-depth centroids in the focal latent subspace ($\mathbf{z}_{\text{focal}}$) across scattering densities, averaged over N = 5 independent training seeds (mean ± std). Inter-cluster distance is the Euclidean distance between cluster centroids in $\mathbb{R}^6$ space.

| | Inter-cluster distance | | |
|---|---|---|---|
| | $f_0 \leftrightarrow f_1$ | $f_0 \leftrightarrow f_2$ | $f_1 \leftrightarrow f_2$ |
| Low Scattering | 2.44±0.15 | 4.43±0.35 | 2.30±0.26 |
| Moderate Scattering | 2.61±0.19 | 4.25±0.22 | 2.52±0.15 |
| High Scattering | 12.41±2.95 | 19.94±4.71 | 7.61±1.75 |

To further characterize the geometric structure underlying this improvement, we computed pairwise inter-cluster distances between the three focal depths' cluster centroids in $\mathbf{z}_{\text{focal}}$. The centroid of each cluster $C_i$ is the mean of all encoded samples belonging to that focal position, and the inter-cluster distance between two clusters is the Euclidean distance between their centroids:

$$d(C_i, C_j) = \| \bar{\mathbf{z}}^{(i)} - \bar{\mathbf{z}}^{(j)} \|_2, \quad \bar{\mathbf{z}}^{(i)} = \frac{1}{|C_i|} \sum_{\mathbf{z}_n \in C_i} \mathbf{z}_n$$

As shown in Table III, inter-cluster distances remain lower in Low and Moderate Scattering, then increase with huge steps in High Scattering, as expected in SS results in Table II. Across all scattering conditions, $f_0 \leftrightarrow f_2$ consistently yields the largest inter-cluster distance, reflecting the physical expectation that the two extreme focal planes produce the most distinct speckle statistics. In Low and Moderate Scattering, the adjacent pairs $f_0 \leftrightarrow f_1$ and $f_1 \leftrightarrow f_2$ are nearly equal, indicating a geometrically symmetric representation. In High Scattering, this symmetry breaks and $f_0 \leftrightarrow f_1$(12.41) becomes notably larger than $f_1 \leftrightarrow f_2$(7.61), suggesting that heavy scattering produces greater speckle divergence between $f_0$ and $f_1$ focal planes than between $f_1$ and $f_2$ planes, which the encoder captures as an asymmetric reorganization of $\mathbf{z}_{\text{focal}}$.

## V. CONCLUSIONS

In this work, we investigated the impact of circular occlusion masking with varying radius on reconstruction quality. We found that the original information still can be found in the peripheral region of images in Scattering datasets compared to No-Scattering dataset that failed drastically as the radius of mask increases. Additionally, we applied curve fitting (Sigmoid for No Scattering dataset and Power law for Scattering datasets), and the resulting parameters effectively illustrated how the Scattering dataset contains more evenly distributed information across the images compared to the No Scattering dataset, with heavier scattering leading to a more uniform distribution.

We also examined whether optical scattering could enable identifying the focal depth of information by projecting the same MNIST pattern onto each of three focal depths of the scattering medium and engineering VAE architecture into split-latent VAE. We discovered an improvement in focal depth classification accuracy as the scattering intensity increased even when high scattering had the lowest SOPCNN image classification accuracy. Furthermore, we calculated silhouette score (SS) and pairwise inter-cluster distances between the three focal latent vector centroids to assess how scattering intensity influences the learned focal latent space. Consequently, we observed that heavy scattering not only actively enhances focal depth discriminability in the learned latent space but also captures and amplifies the asymmetry between focal planes. The focal depth discrimination capability is consistent with the 3D nature of a scattering medium and reveals two competing optimization choices: More scattering elements along a 3D path enable better depth discrimination, but eventually lead to a drop-off in the transmitted information.

*Contact author: wlosert@umd.edu

Overall, this study presents a proof-of-concept that scattering distributes information uniformly across the imaging sensor, and enables AI-assisted focal depth discrimination of optical signals. Our work raises the question whether optical scattering could be used in conjunction with AI-interpretation to develop robust and 3D sensing capabilities in a broad range of science and engineering domains.

## ACKNOWLEDGMENTS

This work was supported by Lockheed Martin University Engagement Funds. We also thank Aiden Heggs and Prof. John Fourkas from University of Maryland, College Park for their help in preparing the scattering media.

*Contact author: wlosert@umd.edu